\documentclass[useAMS,usenatbib]{mn2e}

\usepackage[psamsfonts]{amssymb}
\usepackage[dvips]{graphicx}
\usepackage{amsmath,alltt}
\usepackage{multirow}
\usepackage{rotating}
\usepackage{lscape}

\title[Two-body relaxation modified by gas accretion]{Modifying two-body relaxation in N-body systems by gas accretion}
\author[Leigh, N. W. C., Sills, A. \& B\"{o}ker]{Nathan Leigh$^{1}$, Alison Sills$^{2}$, 
Torsten B\"{o}ker$^{1}$
\thanks{E-mail: nleigh@rssd.esa.int (NL), asills@mcmaster.ca (AS), tboeker@rssd.esa.int (TB)}\\
$^{1}$European Space Agency, Space Science Department, Keplerlaan 1,
2200 AG Noordwijk, The Netherlands \\
$^{2}$McMaster University, Department of Physics and Astronomy, 1280 Main St. W., Hamilton, Ontario, Canada, L8S 4M1}
\begin{document}

\pagerange{\pageref{firstpage}--\pageref{lastpage}} \pubyear{2011}

\maketitle

\label{firstpage}

\begin{abstract}
We consider the effects that accretion from the interstellar medium onto 
the particles of an N-body system has on the rate of two-body relaxation.  To this end, 
we derive an 
accretion-modified relaxation time by adapting Spitzer's two-component model to include 
the damping effects of accretion.  We consider several different mass-dependencies and 
efficiency factors for the accretion rate, as well as different mass ratios for the 
two components of the model.

The net effect of accretion is to accelerate mass segregation by increasing the 
average mass $\bar{m}$, since the relaxation time is inversely proportional to $\bar{m}$.  
Under the assumption that the accretion rate increases with the accretor mass, there are 
two additional effects that accelerate mass segregation.  First, accretion acts to 
increase the range of any initial 
mass spectrum, quickly driving the heaviest members to even higher masses.  Second, 
accretion acts to reduce the velocities of the accretors due to conservation of 
momentum, and it is the heaviest members that are affected the most.  Using our 
two-component model, we quantify these effects as a function of the accretion rate, the 
total cluster mass, and the component masses.  We conclude by discussing the implications 
of our results for the dynamical evolution of primordial globular clusters, primarily in the 
context of black holes formed from the most massive stellar progenitors.
\end{abstract}


\section{Introduction} \label{intro}

For most of the life of a massive star cluster, two-body relaxation is the
dominant physical mechanism driving its evolution
\citep[e.g.][]{henon60, henon73, spitzer87, heggie03, gieles11}.
That is, the cumulative effects of long-range gravitational interactions between 
stars act to alter their orbits within the cluster.  These interactions push the 
cluster toward a state of energy equipartition in which all objects have 
roughly the same kinetic energy.  Consequently, the velocities of the most massive 
objects decrease, and they accumulate in the central regions of the cluster.  Similarly, the 
velocities of the lowest mass objects increase, and they are 
subsequently dispersed to wider orbits.  This mechanism, called mass segregation, also 
contributes to the escape of stars from their host cluster across the tidal boundary, 
with the probability of ejection increasing with decreasing stellar mass.  
Therefore, two-body relaxation acts to slowly modify the radial distribution of stellar
masses within clusters, and can cause very dynamically evolved clusters to
be severely depleted of their low-mass stars \citep[e.g.][]{vonhippel98,demarchi10,leigh12}.

Energy equipartition is an idealized state that should arise after the cumulative 
effects of many long-range interactions.  In 
a real star cluster with a full spectrum of stellar masses, however, equipartition may not 
actually be achievable \citep[e.g.][]{binney87,heggie03}.  
As mentioned, the tendency towards energy equipartition reduces the velocities of the heaviest 
stars, causing them to sink in to the central cluster regions.  Here, they are 
re-accelerated by the central cluster potential and gain kinetic energy.  As this process 
proceeds, it leads to a contraction of the core and subsequently a shorter 
central relaxation time \citep[e.g.][]{spitzer87,heggie03}.  
A shorter relaxation time leads to a faster rate of energy transfer from heavier 
to lighter stars.  Eventually, this makes the heaviest stars evolve 
away from equipartition.

This was first demonstrated by \citet{spitzer69} using analytic techniques 
and a number of simplifying assumptions.  \citet{spitzer69} adopted a 
two-component system with masses $m_{\rm 1}$ and $m_{\rm 2}$ 
(where $m_{\rm 1} > m_{\rm 2}$), forming sub-systems with total masses 
$M_{\rm 1}$ and $M_{\rm 2}$.  Provided that $M_{\rm 1} \ll M_{\rm 2}$, 
Spitzer derived the conditional requirement for a cluster to 
achieve energy equipartition in equilibrium.  Based on this, Spitzer argued 
that energy equipartition could not be achieved in a cluster with a realistic 
mass spectrum, since there should always be enough mass in the heavier species 
for it to form a sub-system in the central cluster regions that decouples 
dynamically from the lighter species.  This is commonly called the 
Spitzer instability \citep[e.g.][]{spitzer87,heggie03,portegieszwart04}.

A particularly compelling example of the Spitzer instability involves 
stellar-mass black holes (BHs) in globular clusters (GCs).  \citet{phinney91} 
first argued that BHs formed from the most massive stars should rapidly 
segregate into the core where they decouple dynamically from the rest of 
the cluster to form a distinct sub-system.  Three-body scattering events then 
lead to the formation of BH-BH binaries, which in turn encounter other BHs and 
BH-BH binaries.  These 3- and 4-body interactions are sufficiently energetic 
to eject the BHs from the cluster.  In the end, most BHs are expected 
to be ejected, leaving only a handful behind.

This picture has recently been challenged in the literature.  In 
particular, several authors have argued that the Spitzer instability 
should break down before most BHs are ejected \citep[e.g.][]{moody09}, and that the time-scale 
for all BHs to be ejected could exceed a Hubble time in some clusters 
\citep[e.g.][]{downing10}.  This view is supported by recent claims 
in the literature that stellar-mass BHs may be present in GCs in surprising 
numbers.  For instance, \citet{strader12} recently reported two flat-spectrum 
radio sources in M22, which appear to be accreting 
stellar-mass BHs.  This suggests that this cluster could contain on the order of 
$\sim 5 - 100$ stellar-mass BHs.  If BHs were indeed efficiently dynamically 
ejected, this, in turn, would suggest that a more substantial population of BHs 
once existed in M22, and likely other GCs as well.

The emerging picture for the formation of massive GCs involves multiple 
episodes of star formation \citep[e.g.][]{piotto07,gratton12,conroy11,conroy12}.  
In this context, \citet{leigh13} recently considered the implications of 
the mass growth of BHs formed from massive progenitors belonging to the 
first generation due to accretion from the interstellar medium.  The 
authors argued that, in principle, BHs could deplete a significant 
fraction of the available gas reservoir within $\lesssim 10^8$ years.  
If BHs were indeed to accrete efficiently from the ISM, they should not only 
grow in mass, but their velocities should also decrease due to conservation 
of momentum.  This should preferentially 
accelerate the process of mass segregation for the BHs, causing them 
to rapidly accumulate in the central regions of the cluster if they 
did not form there in the first place.  This could accelerate the 
dynamical decoupling of the BH sub-population from the rest of the system, and 
hence the phase of dynamical BH ejections due to the Spitzer instability.

In this paper, we consider how accretion from the interstellar 
medium affects the rate of mass segregation in a star cluster.  We are 
especially interested in the implications for BHs in primordial 
GCs.  Thus, 
we re-visit Spitzer's two-component model to derive an accretion-modified 
relaxation time.  We argue that the rate of mass segregation 
should be affected by accretion in the following way.  
First, assuming the accretion rate increases with the 
accretor mass, accretion acts to increase the range of any initial
mass spectrum, driving the heaviest members to higher masses the fastest.  Second,
accretion acts to reduce the velocities of the accretors due to conservation of
momentum, and it is the heaviest members whose velocities are reduced the 
fastest.  Both of these effects exacerbate the Spitzer instability, and 
should accelerate the rate of mass segregation in a primordial star cluster.  

In order to better quantify this qualitative picture, we present our 
adapted version of Spitzer's two-component model in Section~\ref{method}.  
Specifically, we derive an 
accretion-modified relaxation time, as well as the critical accretion rate at 
which the rates of mass segregation due to both two-body relaxation and 
accretion are equal.  We present our results in Section~\ref{results} for 
several different assumptions regarding the total cluster mass and 
accretion rate.  In Section~\ref{discussion}, we discuss the implications of our results 
for both star formation and stellar remnants in primordial globular clusters.  
We summarize our results in Section~\ref{summary}.

\section{Method} \label{method}

In this section, we present our analytic derivation of an accretion-modified 
relaxation time for a two-component model star cluster, and derive the critical
accretion rate required for the mass segregation timescales due to
two-body relaxation and accretion to be equal.  We begin by summarizing briefly 
the relevant background related to both two-body relaxation and accretion.

\subsection{Two-body relaxation} \label{relax}

Consider a two-component model for a star cluster with component masses 
$m_1$ and $m_2$, such that $m_1 > m_2$.  The populations for these two species 
have total masses $M_1$ and $M_2$ with $M_1 \ll M_2$.  We let $v^2$ denote 
the initial mean square speed of \textit{both} species, since at birth the 
cluster is not in a state of energy equipartition.  The e-folding 
time for the tendency to equipartition bears a striking resemblance to the 
relaxation time \citep{heggie03}.  Thus, to order of magnitude, the relaxation 
time can be approximated by calculating the time required for the 
mean square speed of the heavier species to fall from $v^2$ to a value 
$\sim m_2v^2/m_1$.  

If the potential well of the lighter species is modelled using a parabolic 
profile, then equipartition will lead to the heavier species being confined 
to a region of size $\sim r_h\sqrt{m_2/m_1}$ \citep{heggie03}.  
The total mass of the heavier species within this region is $M_1$, whereas that 
for the lighter species is $M_2(m_2/m_1)^{3/2}$.  At this point, however, 
it is not clear whether or not the lighter species remains the dominant mass 
component in this region.  If not, the heavier species becomes increasingly 
affected by its own self-gravity, and can decouple dynamically from the remainder 
of the system.  Consequently, it may only be possible to achieve equipartition 
provided \citep{spitzer69,heggie03}: 
\begin{equation}
\label{eqn:spitzer-criterion}
M_1 \le M_2\Big( \frac{m_2}{m_1} \Big)^{3/2}.
\end{equation}
This is known as Spitzer's criterion.

In general, the half-mass relaxation time approximates the rate of
two-body relaxation throughout the entire
cluster.  In GCs, it ranges from roughly a few hundred million years to the age of the
Universe or longer, and is approximated by \citep{spitzer87}:
\begin{equation}
\label{eqn:t-rh}
\tau_{\rm rh} [yr] = 1.7 \times 10^5[r_{\rm h} [pc]]^{3/2}N^{1/2}[\bar{m} [M_{\odot}]]^{-1/2},
\end{equation}
where $r_{\rm h}$ is once again the half-mass radius, $N$ is the total number of stars
within $r_{\rm h}$, and $\bar{m}$ is the average stellar mass.  We assume that 
the value of $r_{\rm h}$ remains constant in time.  This is reasonable since
simulations have shown that $r_{\rm h}$ changes by not more than a factor of a few 
over the course of a typical cluster's lifetime \citep{henon73,spitzer87,heggie03,webb12}.  
The timescale for mass segregation due to two-body relaxation for an object of 
mass $m$ is then approximately \citep{vishniac78}:
\begin{equation}
\label{eqn:tau-seg-time}
\tau_{\rm seg,2body}(m) = \frac{\bar{m}}{m}\tau_{\rm rh}.
\end{equation}

\subsection{Accretion} \label{accrete}

A strict theoretical upper limit for the accretion rate is given by 
the Bondi-Hoyle limit \citep{bondi44}.  In this approximation, the 
accretion is spherically symmetric, and the forces due to gas 
pressure are insignificant compared to gravitational 
forces.  The background gas is treated as
uniform and either stationary or moving with constant velocity
relative to the accretor.  This assumption gives reasonable accretion
rates in the low-density, low-angular momentum regime.  That is, provided 
the properties of the gas are such that the density, velocity, and total 
angular momentum are low, at least in the vicinity of the accretor, the 
Bondi-Hoyle limit approximately describes the true accretion rate 
\citep[e.g.][]{fryxell88, ruffert94, ruffert97, foglizzo99}.  

For large accretor masses and high gas densities, the Bondi-Hoyle rate 
can become extremely high.  Here, pressure forces could play an important role 
in reducing the accretion rate.  Indeed, this occurs if the outward continuum 
radiation force balances the inward gravitational force.  This limit, called 
Eddington-limited accretion, gives considerably more modest accretion rates 
in the high gas density regime \citep{eddington26,eddington30}.
The Eddington rate should also be much closer to the true 
accretion rate if the gas contains significant angular momentum, and accretion
proceeds mainly via angular momentum re-distribution within a disk \citep[e.g.][]{rybicki79}.

In general, theoretical studies have
shown that the exact accretion rate can deviate substantially from the idealized
cases described by Bondi-Hoyle and Eddington-limited accretion.  For example, when 
an accretor is radiating at above the Eddington luminosity, significant amounts 
of gas can be expelled at high velocities due to 
the intense winds that are initiated \citep[e.g.][]{king03}.  This contributes 
to a reduction in the overall accretion rate.  On the other hand, 
at very high accretion rates, photon-trapping can occur.  This 
makes accretion disks radiatively inefficient and provides a means of circumventing the
Eddington limit \citep{paczynsky80}.  

Numerical studies have also revealed the sensitivity 
of the accretion rate to small-scale gas dynamics.  Recent work by 
\citet{krumholz06} considered gas accretion onto point masses in a 
supersonically turbulent medium characterized by background density
and velocity distributions that vary in both time and space.  The authors
show that in this regime, the accretion rate can either be described by the
classical Bondi-Hoyle approximation, or a vorticity-dominated flow.  In the latter
case, the accretion rate can be significantly reduced relative to what is
predicted by the Bondi-Hoyle limit \citep{krumholz04, krumholz05}. 
Even more recently, \citet{park13} studied the growth and luminosity of BHs in 
motion with respect to their surrounding medium using two-dimensional 
axis-symmetric numerical simulations.  The authors show that the accretion rate 
can actually increase with increasing BH velocity, contrary to the naive 
predictions of simple analytic theory \citep{hoyle39}.  

In summary, theoretical work has shown that a wide range of accretion rates 
are possible.  
We will use the Bondi-Hoyle and Eddington limits for the accretion rate (combined 
with an accretion efficiency parameter) since these provide two different 
dependences on the accretor mass.  However, the derivation presented in the subsequent 
section can be used to model other mass-dependences for the accretion rate as well.

\subsection{Deriving the Relaxation Time} \label{derivation}

We are interested in the mass segregation timescale due to two-body relaxation for 
the heavier species in the two-component model described in Section~\ref{relax}.  To 
first order, this is approximated by:
\begin{equation}
\label{eqn:tau-seg-time}
\tau_{\rm seg,2body}(m_{\rm 1},t) = \frac{\bar{m}(t)}{m_{\rm 1}(t)}\tau_{\rm rh}(t),
\end{equation}
where $\tau_{\rm rh}(t)$ denotes the half-mass relaxation time obtained 
by using the total number of objects $N = N_{\rm 1} + N_{\rm 2}$ and average object 
mass $\bar{m}(t) = (m_{\rm 1}(t)N_{\rm 1} + m_{\rm 2}(t)N_{\rm 2})/(N_{\rm 1} + N_{\rm 2})$ 
in Equation~\ref{eqn:t-rh}.  

We calculate the time-dependence for the mass of an object belonging to species $i$ as 
follows.  First, assuming an 
object of initial mass $m_{\rm i}(0)$ accretes at a rate $\dot{m}_{\rm i} = dm_{\rm i}/dt$ for 
a total time $t$, we have:
\begin{equation}
\label{eqn:mass-time}
m_{\rm i}(t) = m_{\rm i}(0) + \int_{0}^{t} \dot{m}_{\rm i}dt.
\end{equation}
For the accretion rate, we assume a mass-dependence of the form:
\begin{equation}
\label{eqn:acc-rate}
\dot{m}_{\rm i} = {\lambda}{\delta}m_{\rm i}^{\rm \epsilon},
\end{equation}
where $\lambda$, $\delta$, and $\epsilon$ are all free parameters.  The power-law 
exponent $\epsilon$ decides the mass-dependence for the accretion rate.  
The accretion coefficent $\delta$ is derived according to the 
physical assumptions that decide the rate of accretion.  For example, adopting the 
Bondi-Hoyle approximation implies $\epsilon = 2$, and \citep[e.g.][]{bondi44,maccarone12,leigh13}:
\begin{equation}
\label{eqn:mdot-BH}
\delta = 7 \times 10^{-8} {M_{\odot}^{-1}}{\rm yr}^{-1} \Big( \frac{n}{\rm 10^6 cm
^{-3}} \Big) \Big( \frac{\sqrt{c_{\rm s}^2 + v^2}}{\rm 10^6 cm s^{-1}} \Big)^{-3},
\end{equation}
where $n$ is the particle number density, $c_{\rm s}$ is the sound speed, and $v^2$ 
is the root-mean-square speed.  Similarly, assuming Eddington-limited accretion 
implies $\epsilon = 1$, and \citep{eddington26,eddington30}:
\begin{equation}
\begin{gathered}
\label{eqn:mdot-Edd}
\delta = \frac{4{\pi}G}{{\eta}{\kappa}c} \\
       = 2.2 \times 10^{-8} {\rm yr}^{-1},
\end{gathered}
\end{equation}
where $c$ is the speed of light, ${\eta}c^2$
is the accretion yield from unit mass, and ${\kappa}$ is the electron
scattering opacity.  We take $\kappa = 0.34$ cm$^{2}$ g$^{-1}$, which assumes a hydrogen 
mass fraction of $X = 0.7$.  We further adopt $\eta = 0.1$, which is reasonable if the 
accretors are BHs since this parameter quantifies the amount of energy radiated 
away during accretion.  The accretion efficiency 
parameter $\lambda$, on the other hand, is left as a free parameter in our model. 

In order to solve for the accretor mass as a function of time, Equation~\ref{eqn:mass-time} 
can be re-written such that the integral is with respect to mass.  
Provided $\epsilon > 1$, this gives for the mass of an object belonging to 
species $i$ at time $t$:
\begin{equation}
\label{eqn:mass-time2}
m_{\rm i}(t) = \Big( m_{\rm i}(0)^{1-\epsilon} + {\lambda}{\delta}(1-\epsilon)t \Big)^{1/(1-\epsilon)}. 
\end{equation}
Similarly, for $\epsilon = 1$, we have:
\begin{equation}
\label{eqn:mass-time3}
m_{\rm i}(t) = m_{\rm i}(0)e^{\rm {\lambda}{\delta}t}.
\end{equation}
We arrive at the mass segregation timescale due to two-body relaxation for the heavier 
species by substituting either Equation~\ref{eqn:mass-time2} or Equation~\ref{eqn:mass-time3} 
for both species 1 and 2 into Equation~\ref{eqn:tau-seg-time}.  

In addition to two-body relaxation, 
accretion should also affect the stellar velocities via conservation of 
momentum.  Provided the accretion rate increases with increasing accretor mass, this 
will reduce the kinetic energy of the heavier species faster than the lighter 
species, in rough analogy with the effects of two-body relaxation.  Thus, 
accretion-induced changes in the stellar velocities should \textit{accelerate} 
the rate of mass segregation.  We calculate the time needed for 
accretion to change the velocities of the heavier species from a root-mean-square 
speed of $v^2$ to a value of $m_{\rm 2}/m_{\rm 1}v^2$.  
This corresponds to the time for accretion to affect the velocities of the 
heavier species by roughly the same amount as is done by two-body relaxation 
over a single relaxation time.

For a given accretion rate and initial masses, we calculate the time needed 
for the mean square speed of the heavier 
species to reach a value $m_{\rm 2}(t)v^2/m_{\rm 1}(t)$ using conservation of momentum, from 
an initial value $v$.  
The time needed for the heavier species to reach equipartition 
via accretion can then be written:
\begin{equation}
\label{eqn:tau-acc}
\tau_{\rm seg,acc}(m_1,t) = \int_{\rm m_1(t)}^{\rm m_{1,f}(t)}\frac{dm_1}{\dot{m}_1},
\end{equation}
where the final mass is $m_{\rm 1,f} = \sqrt{m_{\rm 1}^3/m_{\rm 2}}$ by conservation 
of momentum.  

Substituting $m_{\rm 1,f}(t) = \sqrt{m_{\rm 1}(t)^3/m_{\rm 2}(t)}$ into 
Equation~\ref{eqn:tau-acc}, we arrive at the timescale needed for accretion 
to push the heavier species into approximate equipartition:
\begin{equation}
\label{eqn:tau-acc2}
\tau_{\rm seg,acc}(m_{\rm 1},t) = \frac{m_{\rm 1}(t)^{1-\epsilon}\Big( (m_{\rm 1}(t)/m_{\rm 2}(t))^{(1-\epsilon)/2} - 1 \Big)}{{\lambda}{\delta}(1 - \epsilon)},
\end{equation}
for $\epsilon > 1$.  Similarly, for $\epsilon = 1$ or Eddington-limited accretion, we 
obtain using Equation~\ref{eqn:mass-time3}:
\begin{equation}
\label{eqn:tau-acc3}
\tau_{\rm seg,acc}(m_{\rm 1},t) = \frac{\ln(m_{\rm 1}(t)/m_{\rm 2}(t))}{2{\lambda}{\delta}}.
\end{equation}
We consider accretion rates with a mass-dependence such that $\epsilon \ge 1$, since this 
includes both Bondi-Hoyle and Eddington-limited accretion, as well as intermediate 
and even steeper mass accretion rates.

The total rate (taken to be the inverse of the total mass segregation timescale 
$\tau_{\rm seg,tot}$) at which the heavier species 
achieves mass segregation can be written as the sum of the rate of two-body 
relaxation and the rate at which accretion pushes the heavier species to
equipartition.  Re-arranging this equation, we arrive at the total 
accretion-modified mass segregation timescale for the heavier species:
\begin{equation}
\label{eqn:t-eq}
\tau_{\rm seg,tot}(m_{\rm 1},t) = \frac{\tau_{\rm seg,acc}(m_{\rm 1},t)\tau_{\rm seg,2body}(m_{\rm 1},t)}{\tau_{\rm seg,acc}(m_{\rm 1},t) + \tau_{\rm seg,2body}(m_{\rm 1},t)}.
\end{equation}

\subsection{Deriving the critical accretion rate} \label{critical}

To derive the critical accretion rate $\delta_{\rm crit}$ at which two-body 
relaxation and accretion drive the mass segregation process at the same rate, 
we set $\tau_{\rm seg,acc} = \tau_{\rm seg,2body}$ and solve for $\delta$ as 
a function of $\epsilon$, $\lambda$, $m_{\rm 1}$, $m_{\rm 2}$, and 
$t_{\rm rh}$.  This gives for $\epsilon > 1$:
\begin{equation}
\label{eqn:delta2}
\delta_{\rm crit} = \frac{m_{\rm 1}(t)^{2-\epsilon}\Big( (m_{\rm 1}(t)/m_{\rm 2}(t))^{(1-\epsilon)/2} - 1 \Big)}{{\lambda}(1 - \epsilon)\bar{m}(t)t_{\rm rh}(t)}.
\end{equation}

The procedure is similar for $\epsilon = 1$, except we use Equation~\ref{eqn:tau-acc3} 
instead of Equation~\ref{eqn:tau-acc2}.  This gives for Eddington-limited accretion:
\begin{equation}
\label{eqn:delta3}
\delta_{\rm crit} = \frac{m_{\rm 1}(t)\ln(m_{\rm 1}(t)/m_{\rm 2}(t))}{2{\lambda}\bar{m}(t)t_{\rm rh}(t)}.
\end{equation}

In Section~\ref{results}, we will use Equation~\ref{eqn:delta2} and Equation~\ref{eqn:delta3} 
in order to study the interplay between our assumptions regarding the gas properties, 
which affect the accretion rate, and our assumption for the total cluster mass, which 
determines the rate of two-body relaxation.

\subsection{Accretion efficiency} \label{lambda-time}

Given our limited understanding of the precise physics of accretion onto
a BH, it is not possible to reliably define a functional form for the accretion 
efficiency parameter $\lambda$. In principle, any realistic accretion model 
should include a \textit{time-dependence} for $\lambda(t)$.\footnote{Alternatively, 
the time-dependence can be absorbed directly into the parameter $\delta$ in 
Equations~\ref{eqn:tau-acc2} and~\ref{eqn:tau-acc3}.}  
For example, fluctuations in the local gas density due to turbulence, a gradual 
or even sudden depletion of the available gas reservoir, or dynamical interactions 
between accreting objects may cause the accretion efficiency to vary over time. 

Our analysis is easily modified to treat time-dependent accretion rates 
by substituting an appropriate choice for $\lambda(t)$ into either Equation~\ref{eqn:tau-acc2} 
or~\ref{eqn:tau-acc3}, and then solving for $\tau_{\rm seg,acc}$. Plausible choices for $\lambda(t)$ 
may either oscillate or decline (steadily or abruptly) in time. The first case, i.e. an
oscillating accretion efficiency, is more easily understood, because under these circumstances, 
our analysis can simply be interpreted as discussing the \textit{time-averaged} accretion 
efficiency parameter.  Thus, in the subsequent sections, we assume an oscillating (or constant) 
accretion efficiency parameter, and discuss only the time-averaged value.

For example, the function $\lambda(t) = (1+{\rm sin}({\pi}t/t_{\rm 0}))$ oscillates between 
0 and 2 with a frequency of $2/t_{\rm 0}$.  In this case, the time-averaged value for $\lambda(t)$ 
is equal to 1, so that 
the time-averaged value for $\tau_{\rm seg,acc}$ remains 
the same as for a constant $\lambda = 1$. 

Accretion efficiency parameters that oscillate in time should be suitable to cases 
where the accretors have alternating ``on'' and ``off'' phases.  This may well be the case 
with accreting BHs, since the radiation emitted due to accretion can heat the 
surrounding gas, which in turn decreases the accretion rate \citep[e.g.][]{blaes95}.  In this case, 
the source of energetic photons responsible for heating the gas is 
turned off, allowing the gas to cool and accretion to re-start in a 
``feedback regulated'' loop \citep[e.g.][]{king03,yuan09}.  

Accretion efficiency parameters that decline in time should be 
appropriate to cases where the available gas reservoir is depleted 
over time.  This could arise gradually if the gas is used to form stars, 
or if significant quantities of gas are accreted by BHs.  Alternatively, the gas 
reservoir could be depleted suddenly, e.g. due to, energy injected from 
supernovae, stellar winds, or winds from accreting compact objects.  In either 
case, Equations~\ref{eqn:tau-acc2} and~\ref{eqn:tau-acc3} should include 
the explicit time-dependence for the accretion efficiency parameter.  This 
will contribute to an increase in $\tau_{\rm seg,acc}$ with time, since the 
decreasing gas mass should translate into a decreasing gas density, and hence 
accretion rate.  In Section~\ref{results}, we will assume that the amount of 
gas lost from the system is negligible over the calculated mass segregation 
timescales, and our interpretation of $\lambda$ as a time-independent quantity 
remains valid.  This is reasonable provided the mass segregation 
timescales due to accretion are much less than the timescale for gas 
depletion.  As we will show in Section~\ref{results}, the current picture 
for the formation of globular clusters and their multiple populations is 
consistent with this scenario \citep[e.g.][]{krause12,krause13,leigh13}. 

\section{Results} \label{results}

In this section, we present the results of our analytic two-component 
model for an accretion-modified two-body relaxation time.   Our aim 
is to quantify the relative rates at which two-body relaxation and 
gas accretion drive a star cluster towards mass segregation, as a function of 
our assumptions for the gas properties, component masses, and total system mass.  
To this end, we present the time evolution of all three mass segregation 
timescales, namely $\tau_{\rm seg,acc}$, 
$\tau_{\rm seg,2body}$, and $\tau_{\rm seg,tot}$, and discuss the critical 
accretion rate required for the mass segregation timescales due to 
two-body relaxation and accretion to be equal as a function of the 
mass-dependence for the accretion rate.

\subsection{Time evolution of the mass segregation timescales} \label{time-evolution}

We begin by quantifying the relative rates of mass segregation due to two-body 
relaxation and gas accretion for different model assumptions.  Specifically, 
we consider several different mass ratios and total system masses for our 
two-component model, as well as different mass-dependences for the rate of 
accretion.  This is meant to quantify the sensitivity of the two different mass 
segregation mechanisms to the cluster and gas properties that decide their rates.

First, we describe our assumptions for the two-component model star cluster, which 
are needed in order to calculate $\tau_{\rm seg,2body}$.  We adopt 
$m_{\rm 2} = 1$ M$_{\odot}$ for the lighter species, but consider two different 
masses for the heavier species, namely $m_{\rm 1} = 10$ M$_{\odot}$ 
and $m_{\rm 1} = 50$ M$_{\odot}$.  We assume a population size of $N_{\rm 1} = 10^2$ for 
the heavier species, but vary the population size of the lighter species by considering 
the values $N_{\rm 2} = 10^5, 10^6, 10^7$.  The component masses and population sizes 
are chosen to represent 
reasonable mass ratios between the average stellar and BH masses, and to ensure 
that the Spitzer criterion (i.e. Equation~\ref{eqn:spitzer-criterion}) 
is initially satisfied.  We adopt a half-mass radius for our model cluster of 
$r_{\rm h} = 10$ pc, and note that assuming a lower value for the half-mass radius would 
only shorten the calculated mass segregation timescales.  

Next, we describe our assumptions for the properties of the accreted gas, which are 
needed to calculate $\tau_{\rm seg,acc}$, and therefore $\tau_{\rm seg,tot}$.  We 
assume a uniform time-independent gas 
density throughout the cluster, so that the accretion rate changes only with
the stellar mass.  We further assume that the gas is always at rest
relative to the accretor when calculating the final accretor velocity 
using conservation of momentum.  For the accretors, we adopt a root
mean-square-speed of $v = 10$ km s$^{-1}$, which is guided by the relation 
$v = \sqrt{2GM/5r_h}$ \citep{binney87} for a total cluster mass 
$M \sim 10^5-10^6$ M$_{\odot}$.  For the gas, we assume a sound speed of
$c_{\rm s} = 10$ km s$^{-1}$, and a particle number density of $n = 10^6$ cm$^{-3}$.  
These assumptions should be 
reasonable for what is expected in a massive primordial GC for the first $\sim 10^8$ years 
\citep[e.g.][]{dercole08,maccarone12,conroy12,krause12,krause13,leigh13}.

\begin{figure*}
\centering
\resizebox{0.46\hsize}{!}{\rotatebox{0}{\includegraphics{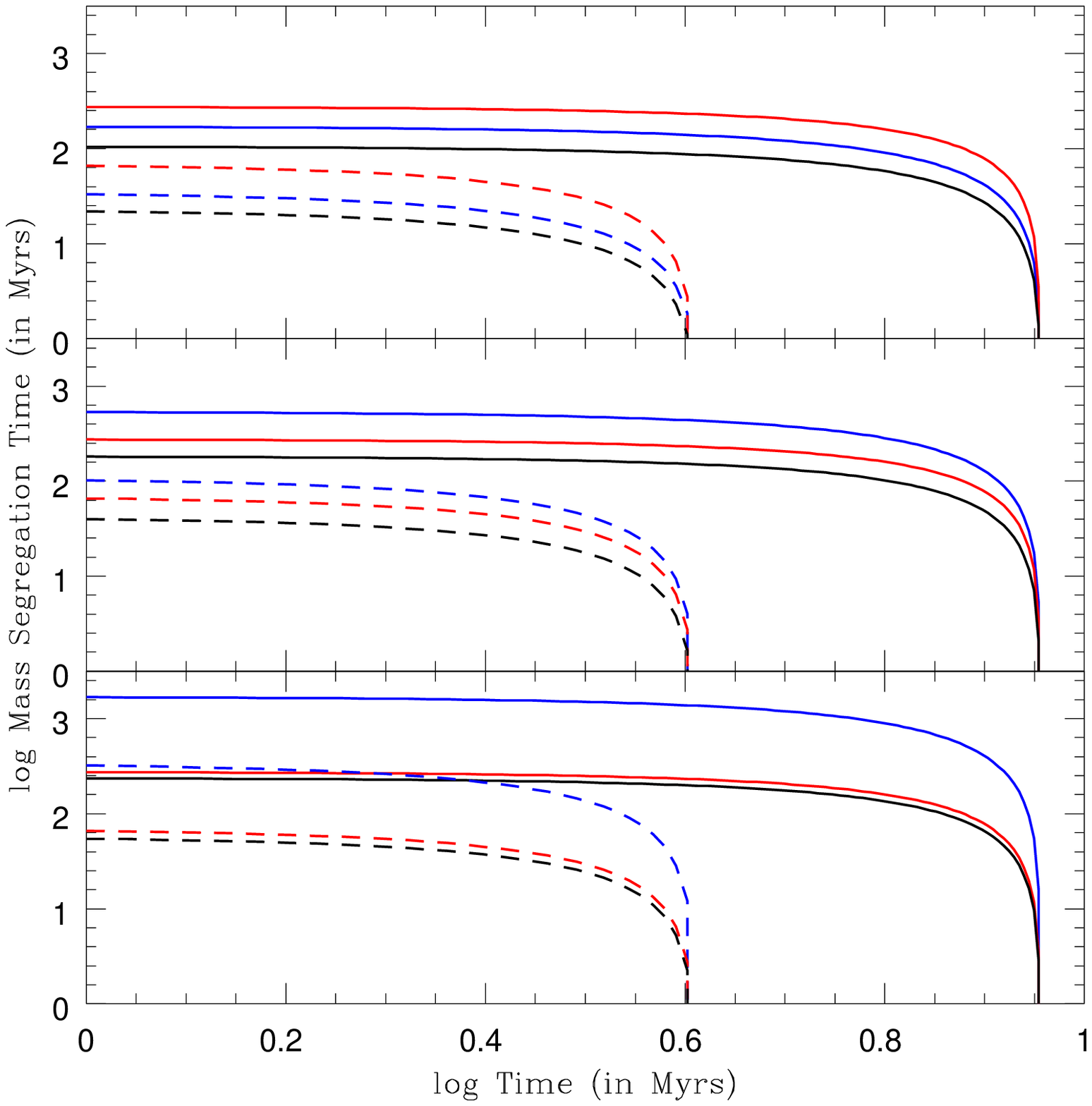}}}
\resizebox{0.46\hsize}{!}{\rotatebox{0}{\includegraphics{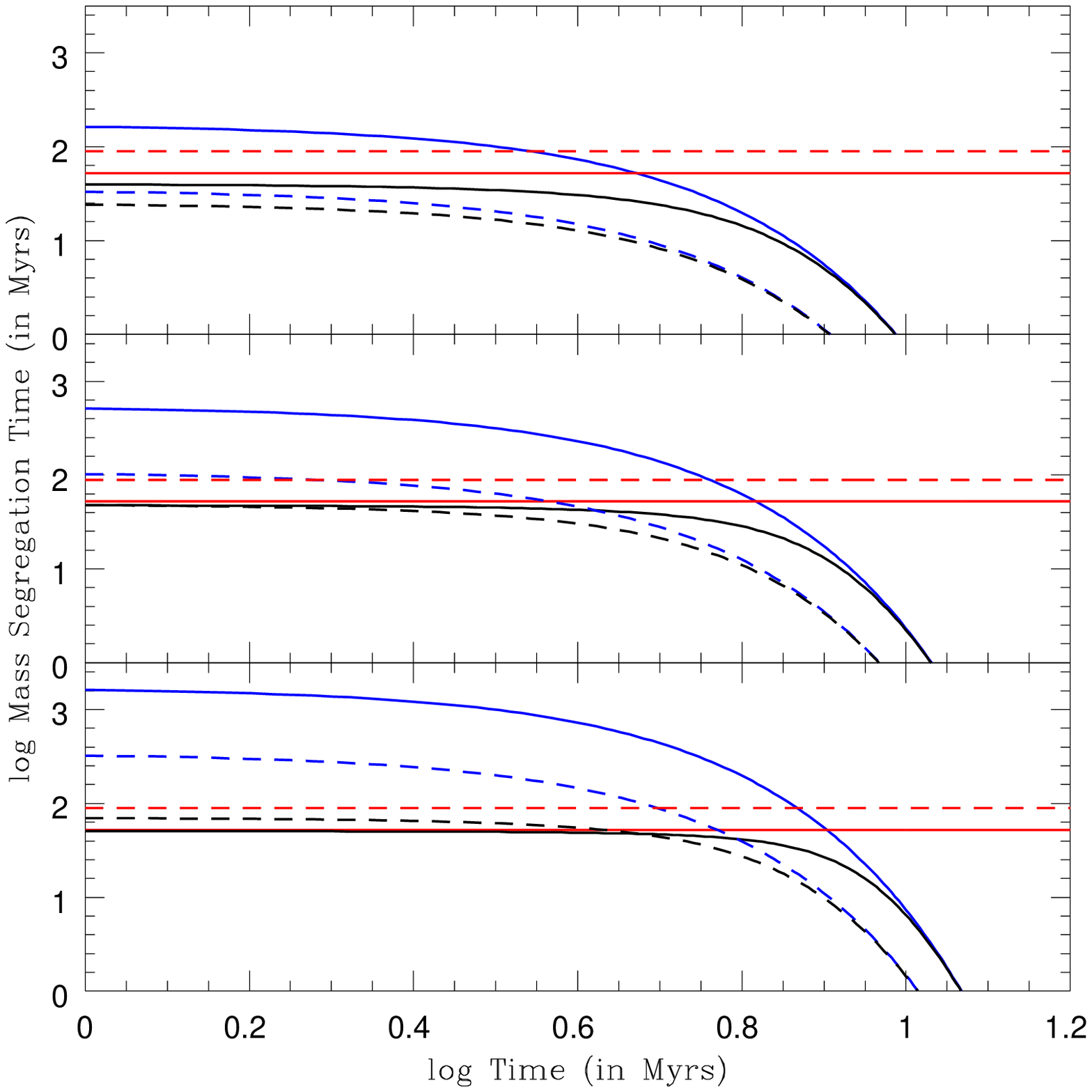}}}
\caption[The mass segregation timescales due to two-body relaxation, accretion, and 
the sum of their rates assuming Bondi-Hoyle (left) and Eddington-limited (right) 
accretion]{The mass segregation timescales 
due to two-body relaxation (blue curves), accretion (red curves), and the combined 
effects of both (black curves), assuming either the Eddington limit with $\lambda = 1.0$ 
for the accretion efficiency parameter (right panels), or the 
Bondi-Hoyle limit with $\lambda = 0.1$ (left panels).  The solid 
and dashed curves apply to populations of 10 M$_{\odot}$ and 50 M$_{\odot}$ black 
holes, respectively.  The top, middle, and bottom insets correspond to total 
population sizes for the lighter species of $N_{\rm 2} = 10^5$, $10^6$, $10^7$, 
respectively.
\label{fig:tau-seg-BH}}
\end{figure*}

We show our results for two different mass-dependencies for the accretion rate.  
The left panels in Figure~\ref{fig:tau-seg-BH} show our results assuming $\epsilon = 2$ in 
Equation~\ref{eqn:acc-rate}, which corresponds to Bondi-Hoyle accretion.  We use 
Equation~\ref{eqn:mdot-BH} for $\delta$, and 
$\lambda = 0.1$ for the accretion efficiency parameter.  
The panels to the right in Figure~\ref{fig:tau-seg-BH} show our results assuming 
Eddington-limited accretion, which means that $\epsilon = 1$ in 
Equation~\ref{eqn:acc-rate} and we use Equation~\ref{eqn:mdot-Edd} for $\delta$.  

The main conclusion to be drawn from Figure~\ref{fig:tau-seg-BH} 
is that, for all but the least massive clusters 
and the lowest accretion rates considered here, the rate of mass segregation due 
to accretion can actually exceed the rate due to two-body relaxation.  The timescale 
at which this occurs is on the order of $\sim 10^8$ years.  Interestingly, this timescale 
is similar to the total time thought to be required for multiple episodes 
of star formation to occur in primordial GCs \citep[e.g.][]{conroy11,conroy12}.  
Thus, our results suggest that accretion from the ISM could significantly affect 
both the spatial and velocity distributions of the heaviest objects in a primordial 
GC \textit{before} the gas reservoir is depleted.  For a typical primordial 
GC, this should be the case provided 
the average accretion rate is greater than $\sim 5-10$\% of the 
Eddington-limited rate, assuming the mass-dependence for the accretion 
rate is linear.  Similarly, if the accretion rate scales with the square of the accretor 
mass, then accretion from the ISM is non-negligible as long as the average accretion
rate is greater than 1-10\% of the Bondi-Hoyle rate.

\subsection{The critical accretion rate} \label{critical2}

In this section, we calculate the critical accretion rate required for the 
rates of mass segregation due to two-body relaxation and gas accretion to be equal.  
Our aim is to quantify the relative importance of the different parameters for 
the cluster and gas properties in establishing a balance between the competing 
effects of two-body relaxation and accretion.

\begin{figure}
\begin{center}
\includegraphics[width=\columnwidth]{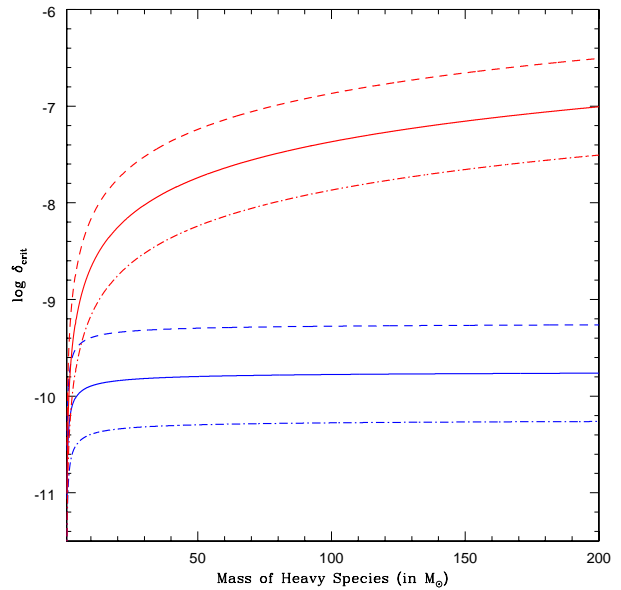}
\end{center}
\caption[The critical accretion rate at which both mass segregation timescales are 
equal shown as a function of the accretor mass]{The critical accretion rate $\delta_{\rm crit}$ 
at which the mass segregation timescales due to two-body relaxation and accretion 
are equal, shown as a function of the mass of the heavier species $m_{\rm 1}$.  The 
blue curves show the results assuming Bondi-Hoyle accretion, whereas the red 
curves are for Eddington-limited accretion.  The dashed, solid, and dash-dotted lines 
correspond to $N_{\rm 1} = 10^5, 10^6, 10^7$, respectively.  For both sets of curves, we 
assume a constant mass for the lighter species of $m_{\rm 2} = 1$ M$_{\odot}$, and 
we set $\lambda = 1.0$ for the accretion efficiency parameter.
\label{fig:delta-crit}}
\end{figure}

In Figure~\ref{fig:delta-crit}, we show the critical accretion rate $\delta_{\rm crit}$ 
as a function of the mass of the heavier species $m_{\rm 1}$.  These results are 
calculated using Equations~\ref{eqn:delta2} and \ref{eqn:delta3}, which 
correspond to Bondi-Hoyle (blue) and Eddington-limited (red) 
accretion, respectively.  In both cases, we assume a constant mass for the 
lighter species of $m_{\rm 2} = 1$ M$_{\odot}$, a constant population 
size for the heavier species of $N_{\rm 1} = 10^2$, and a constant accretion 
efficiency parameter $\lambda = 1.0$.  In order to vary 
the rate of two-body relaxation without affecting the rate of mass segregation 
due to accretion, we consider three different population sizes 
for the lighter species, namely $N_{\rm 2} = 10^5, 10^6, 10^7$.  

Figure~\ref{fig:delta-crit} shows that for the case of Eddington-limited accretion, the 
critical accretion rate increases with increasing accretor mass.  This is because, as the 
accretor mass increases, the mass segregation timescale due to two-body relaxation 
decreases faster than the mass segregation timescale due to accretion.  In the case of 
Bondi-Hoyle accretion, however, the critical accretion rate depends only 
very weakly on the accretor mass, which is due to the fact that the accretion rate 
scales as the square of the accretor mass.  We emphasize that with the exception of the 
Eddington-limited rate at 
large accretor masses, the critical accretion rates are comparable to, or even smaller 
than, those observed in nearby star-forming regions \citep[e.g.][]{mckee07}.

\section{Discussion} \label{discussion}

In this section, we discuss the implications of our results for mass segregation 
in primordial globular clusters, in particular with regards to black holes.

\subsection{Enhanced mass segregation} \label{primordial}

One of the key conclusions arising from our analysis is that accretion should 
accelerate the rate at which a star cluster becomes mass 
segregated compared to two-body relaxation alone.  In fact, accretion can dominate over 
two-body relaxation in massive clusters for accretion rates 
that are below the Bondi-Hoyle or Eddington-limited rates by one or even 
two orders of magnitude.  This is because the relaxation time 
increases with the cluster mass, whereas the mass segregation timescale due to 
accretion is independent of the cluster mass (assuming that the gas properties 
are independent of the cluster mass).  Our results suggest that two-body relaxation 
should dominate the mass segregation process in low-mass primordial clusters with global 
relaxation times $\lesssim 10^7-10^8$ years and hence total cluster masses 
$\lesssim 10^4-10^5$ M$_{\odot}$, provided that our models assumptions are valid.  In 
this regime, accretion should only have a small effect on the rate of mass segregation, 
and long-range gravitational 
interactions should alter the accretors' velocities faster than they are 
reduced by the accretion process.  In more massive clusters, however, the damping effects of 
accretion could play a significant role in accelerating the rate of mass segregation.  

\subsection{The effects of a realistic mass spectrum} \label{mass-spec}

Our assumption of a two-component model serves to demonstrate the 
effects of accretion from the ISM on a cluster's dynamical evolution.  
The qualitative nature of our results should hold if a realistic mass spectrum is 
adopted instead.  Accretion can modify the distribution of 
velocities on relatively short timescales in gas-embedded clusters.  How exactly 
the velocities become modified depends on several parameters, in particular the 
mass spectrum, the total cluster mass, the properties of the gas, and the 
accretion rate.  

In general, we expect accretion to amplify or exacerbate the Spitzer 
instability.  This is due to the mass-dependence of the accretion rate, and the fact that 
typically $\epsilon > 0$ in Equation~\ref{eqn:acc-rate} (i.e. the accretion rate), which 
causes the more massive component to grow in mass the fastest.  Thus, according to 
Equation~\ref{eqn:spitzer-criterion}, Spitzer's criterion should typically break down 
sooner as a result of accretion.  However, accretion also acts to reduce the velocities 
of the accretors due to conservation of momentum, and this should most strongly impact 
the most massive objects due once again to the mass-dependence of the accretion rate.  This 
can actually serve to combat the effects of the Spitzer instability by inhibiting the 
most massive objects from decoupling dynamically from the rest of the system once they 
have segregated to the central regions of the cluster.  Clearly, a more sophisticated 
treatment will be needed in future studies in order to properly quantify these effects 
and their implications for the Spitzer instability, and a cluster's ability to 
achieve energy equipartition.

\subsection{Gas properties and the accretion rate} \label{gas-prop}

We stress that our results depend sensitively on our assumption for the accretion rate,
which is poorly constrained, both theoretically and observationally.  
Indeed, the accretion efficiency parameter $\lambda$ adopted in Equation~\ref{eqn:acc-rate} 
is needed to account for the many sources of uncertainty in the gas properties, 
and hence the accretion rate.  
For example, our assumption of a uniform, time-independent gas density is 
an over-simplification.  For one, stellar winds and supernovae could create 
over- and under-densities in the form of sheets and/or filaments, and the efficiency 
of these processes should fluctuate in time given the presence of a realistic mass 
function combined with stellar evolution and the cluster dynamics 
\citep[e.g.][]{krause12,krause13}.  These effects could contribute to a 
reduction in the accretion rate by increasing the relative velocity between the gas 
and the accretors, or by reducing the gas density along the trajectories of the 
accretors.  Realistic hydrodynamical simulations of star cluster formation 
will be needed in order to properly quantify these effects and their 
implications for the accretion rate.  

Our results can be used to guide the parameter space relevant to these future studies.  
In particular, we have placed a 
lower limit on the minimum accretion rate required for accretion to 
significantly affect the distribution of stellar velocities on timescales shorter than 
the relaxation time, as a function of the cluster and gas properties.  Specifically, the 
results of our simple model suggest that, for a typical primordial GC, 
the average accretion rate cannot be much less than $\sim 5-10$\% of the 
Eddington-limited rate, assuming the mass-dependence for the accretion 
rate is linear.  Similarly, for our model assumptions, the average accretion rate cannot 
be much less than $1-10$\% of the Bondi-Hoyle rate if the accretion rate scales 
with the square of the accretor mass.

We have adopted the same root-mean-square speed for all models, independent 
of the total cluster mass.  This is a reasonable assumption since the 
root-mean-square speed scales 
as $v \propto (M/r_{\rm h})^{0.5}$, and $r_{\rm h}$ itself depends weakly 
on the total cluster mass.  Thus, in total, the root-mean-square speed depends 
only very weakly 
on the total cluster mass.  Nevertheless, if the accretion rate scales 
inversely with the velocity of the accretor, as is the case with the 
Bondi-Hoyle approximation, then the dependence of the root-mean-square speed 
on the total cluster mass should contribute to a decrease in the accretion rate 
with increasing cluster mass.  A proper treatment of this effect is beyond 
the scope of this paper, however, it should certainly be considered in 
future studies.

\subsection{Black hole dynamics} \label{BH-dynamics}

The results presented in this paper are especially relevant for black holes 
in primordial globular clusters, since they should be the most massive objects 
in the cluster within a few Myrs of its formation.  Recent evidence suggests 
that there should be a substantial gas reservoir in GCs for the first 
$\sim 10^8$ years \citep[e.g.][]{conroy11}, 
albeit perhaps intermittently, and that nearly all BHs should form from the most massive 
cluster members within the first few Myrs \citep[e.g.][]{maeder09}.  It follows that 
the BHs could have on the order of $10^8$ years to accrete gas from the ISM.  
Additionally, since any BHs formed from progenitors more massive than 
$\sim 50$ M$_{\odot}$ are only slightly less massive than the progenitors 
themselves and do not experience natal kicks \citep{fryer12}, these BHs should 
have both the shortest mass segregation timescales due to two-body relaxation and 
the highest accretion rates (ignoring BH winds and/or Compton heating; see below).  

The key point is that accretion should act to reduce the mass segregation times of 
BHs in primordial GCs, and that this effect could be dramatic.  Beyond this, more 
detailed modeling will be needed to determine the fates of the BHs.  In 
particular, should the increased rate of mass segregation contribute to 
accelerating the onset of the hypothesized phase of dynamical BH ejections?  Or 
could the damping effects of accretion be so dramatic that the BHs are driven to 
merge \citep[e.g.][]{davies11}?  
If so, the formation of an intermediate-mass BH (IMBH) could be the inevitable 
result.  Alternatively, it could be that black hole winds are sufficiently powerful to 
eject the bulk of the gas from the cluster.  Another possibility is that 
the gas in the immediate vicinity of the BHs becomes very hot due to, for 
example, Compton heating \citep{blaes95,yuan09}, such that the accretion rate becomes 
drastically reduced and BH growth is severely limited?

A better understanding of 
how the presence of significant quantities of gas modifies the black hole dynamics in a 
primordial GC could help to constrain the initial cluster conditions.  For example, if 
massive BHs should inevitably merge in the presence of gas but no IMBHs are observed in 
present-day GCs, does this necessarily imply that the BHs never formed in the first place?  
If so, this would suggest that stars with masses $\gtrsim 50$ M$_{\odot}$ must have been 
rare.  This could be the case, for instance, if massive 
primordial GCs were assembled from the mergers of many low-mass 
sub-clumps, as opposed to a single monolothic collapse.  This is because the mass 
of the most massive cluster member correlates with the total cluster mass 
\citep[e.g.][]{kirk11,kirk12}, and hence the massive stellar 
progenitors of the most massive BHs are unlikely to form in low-mass clusters.

\section{Summary} \label{summary}

In this paper, we have considered the effects of accretion from the interstellar medium on the 
rate of two-body relaxation in a star cluster.  To do this, we derived an 
accretion-modified relaxation time by adapting Spitzer's two-component model to include
the effects of accretion.  We considered several different mass-dependencies and
efficiency factors for the accretion rate, as well as different mass ratios for the
two components of the model.

We have shown that accretion acts to increase the rate of mass segregation.  
This is because the relaxation time is inversely proportional to the average 
mass, which increases due to accretion.  There are two additional effects that 
accelerate the mass segregation process, assuming that the accretion rate increases 
with the accretor mass.  First, accretion acts to increase the range of any initial 
mass spectrum, quickly driving the heaviest members to even higher masses.  Second,
accretion acts to reduce the velocities of the accretors due to conservation of
momentum, and it is the heaviest members that are affected the most.  Using our 
two-component model, these effects have been quantified as a function of the accretion rate, the
total cluster mass, and the component masses.  We have discussed our results in the context 
of the dynamical evolution of primordial globular clusters and their black hole sub-populations. 

\section*{Acknowledgments}

We kindly thank an anonymous referee for helpful suggestions that improved our manuscript, as 
well as Cole Miller for useful discussions.  AS is supported by NSERC.


\bsp

\label{lastpage}


\begin{thebibliography}{99}

\bibitem[\protect\citeauthoryear{Baumgardt \& Makino}{2003}]{baumgardt03}
Baumgardt H., Makino J. 2003, MNRAS, 340, 227
\bibitem[\protect\citeauthoryear{Baumgardt, De Marchi \&
    Kroupa}{2008}]{baumgardt08} Baumgardt H., De Marchi G., Kroupa
  P. 2008, ApJ, 685, 247
\bibitem[\protect\citeauthoryear{Binney \& Tremaine}{1987}]{binney87}
  Binney J., Tremaine S., 1987, Galactic Dynamics (Princeton:
  Princeton University Press)
\bibitem[\protect\citeauthoryear{Blaes, Warren \& Madau}{1995}]{blaes95} Blaes O., 
Warren O., Madau P. 1995, ApJ, 454, 370
\bibitem[\protect\citeauthoryear{Bondi \& Hoyle}{1944}]{bondi44} Bondi H.,
Hoyle F. 1944, MNRAS, 104, 273
\bibitem[\protect\citeauthoryear{Conroy \& Spergel}{2011}]{conroy11} Conroy C., Spergel D. N.
2011, ApJ, 726, 36
\bibitem[\protect\citeauthoryear{Conroy}{2012}]{conroy12} Conroy C. 2012, ApJ, 758, 21
\bibitem[\protect\citeauthoryear{Davies, Miller \& Bellovary}{2011}]{davies11} Davies M. B., 
Miller M. C., Bellovary J. M. 2011, ApJ, 740, 42
\bibitem[\protect\citeauthoryear{De Angeli et al.}{2005}]{deangeli05}
  De Angeli F., Piotto G., Cassisi S., Busso G., Recio-Blanco A.,
  Salaris M., Aparicio A., Rosenberg A. 2005, AJ, 130, 116
\bibitem[\protect\citeauthoryear{De Marchi, Paresce \&
    Pulone}{2007}]{demarchi07} De Marchi G., Paresce F., Pulone
  L. 2007, ApJ, 656, L65
\bibitem[\protect\citeauthoryear{De Marchi, Paresce \& Portegies
    Zwart}{2010}]{demarchi10} De Marchi G., Paresce F., Portegies
  Zwart S. 2010, ApJ, 718, 105
\bibitem[\protect\citeauthoryear{D'Ercole et al.}{2008}]{dercole08} D'Ercole A.,
Vesperini E., D'Antona F., McMillan S. L. W., Recchi S. 2008, MNRAS, 391, 825
\bibitem[\protect\citeauthoryear{Downing et al.}{2010}]{downing10} Downing J. M. B., 
Benacquista M. J., Giersz M., Spurzem R. 2010, MNRAS, 407, 1946
\bibitem[\protect\citeauthoryear{Dopita \& Smith}{1986}]{dopita86} Dopita M. A.,
Smith G. H. 1986, ApJ, 304, 283
\bibitem[\protect\citeauthoryear{Eddington}{1926}]{eddington26} Eddington A. S. 1926, The
Internal Constitution of the Stars (Cambridge:  Cambridge University Press)
\bibitem[\protect\citeauthoryear{Eddington}{1930}]{eddington30} Eddington A. S. 1930,
MNRAS, 90, 279
\bibitem[\protect\citeauthoryear{Fall \& Zhang}{2001}]{fall01} Fall
  S. M., Zhang Q. 2001, ApJ, 561, 751
\bibitem[\protect\citeauthoryear{Foglizzo \& Ruffert}{1999}]{foglizzo99} Foglizzo T.,
Ruffert M. 1999, A\&A, 347, 901
\bibitem[\protect\citeauthoryear{Fryer \& Kalogera}{2001}]{fryer01} Fryer C. L.,
Kalogera V. 2001, ApJ, 554, 548
\bibitem[\protect\citeauthoryear{Fryer et al.}{2012}]{fryer12} Fryer C. L.,
Belczynski K., Wiktorowicz G., Dominik M., Kalogera V., Holz D. E. 2012, ApJ, 749, 91
\bibitem[\protect\citeauthoryear{Fryxell \& Taam}{1988}]{fryxell88} Fryxell B. A.,
Taam R. E. 1988, ApJ, 335, 862
\bibitem[\protect\citeauthoryear{Gieles, Heggie \&
    Zhao}{2011}]{gieles11} Gieles M., Heggie D., Zhao H. 2011, MNRAS,
  accepted
\bibitem[\protect\citeauthoryear{Gratton, Carretta \& Bragaglia}{2012}]{gratton12}
Gratton R., Carretta E., Bragaglia A. 2012, Astronomy \& Astrophysics Review,
in press (arXiv:1201.6526)
\bibitem[\protect\citeauthoryear{Harris}{1996, 2010 update}]{harris96}
  Harris, W. E. 1996, AJ, 112, 1487 (2010 update)
\bibitem[\protect\citeauthoryear{Heggie \& Hut}{2003}]{heggie03}
  Heggie D. C., Hut P. 2003, The Gravitational Million-Body Problem:
  A Multidisciplinary Approach to Star Cluster Dynamics (Cambridge:
  Cambridge University Press)
\bibitem[\protect\citeauthoryear{Heggie \& Giersz}{2008}]{heggie08} Heggie D. C.,
Giersz M. 2008, MNRAS, 389, 1858
\bibitem[\protect\citeauthoryear{Heggie \& Giersz}{2009}]{heggie09} Heggie D. C.,
  Giersz M. 2009, MNRAS, 397, 46
\bibitem[\protect\citeauthoryear{Henon}{1960}]{henon60} Henon M. 1960,
  Annales d'Astrophysique, 23, 668
\bibitem[\protect\citeauthoryear{Henon}{1973}]{henon73} Henon
  M. 1973, Dynamical Structure and Evolution of Dense Stellar Systems,
  ed. L. Martinet \& M. Mayor (Geneva Obs.)
Pringle J. E. 2006, MNRAS, 373, L90
\bibitem[\protect\citeauthoryear{Hoyle \& Lyttleton}{1939}]{hoyle39} Hoyle F., 
Lyttleton R. A. 1939, in Proceedings of the Cambridge Philosophical Society, 35
\bibitem[\protect\citeauthoryear{King \& Pounds}{2003}]{king03} King A. R.,
Pounds K. A. 2003, MNRAS, 345, 657
\bibitem[\protect\citeauthoryear{Kirk \& Myers}{2011}]{kirk11} Kirk H., Myers P. C. 2011,
ApJ, 727, 64
\bibitem[\protect\citeauthoryear{Kirk \& Myers}{2012}]{kirk12} Kirk H., Myers P. C. 2012,
ApJ, 745, 131
\bibitem[\protect\citeauthoryear{Krause et al.}{2012}]{krause12} Krause M., Charbonnel C.,
Decressin T., Meynet G., Prantzos N., Diehl R. 2012, A\&A, 546, L5
\bibitem[\protect\citeauthoryear{Krause et al.}{2013}]{krause13} Krause M., Charbonnel C.,
Decressin T., Meynet G., Prantzos N.. 2013, A\&A, accepted (arXiv:1302.2494)
\bibitem[\protect\citeauthoryear{Krumholz, McKee \& Klein}{2004}]{krumholz04} Krumholz M. R.,
McKee C. F., Klein R. I. 2004, ApJ, 611, 399
\bibitem[\protect\citeauthoryear{Krumholz, McKee \& Klein}{2005}]{krumholz05} Krumholz M. R.,
McKee C. F., Klein R. I. 2005, ApJ, 618, 757
\bibitem[\protect\citeauthoryear{Krumholz, McKee \& Klein}{2006}]{krumholz06} Krumholz M. R.,
McKee C. F., Klein R. I. 2006, ApJ, 638, 369
\bibitem[\protect\citeauthoryear{Leigh et al.}{2012}]{leigh12} Leigh N. W.,
Umbreit S., Sills A., Knigge C., Glebbeek E., Sarajedini A. 2012, MNRAS, 422, 1592 
\bibitem[\protect\citeauthoryear{Leigh et al.}{2013}]{leigh13} Leigh N. W.,
B\"{o}ker T., Maccarone T. J., Perets H. B. 2013, MNRAS, 429, 2997
\bibitem[\protect\citeauthoryear{Maccarone \& Zurek}{2012}]{maccarone12}
Maccarone T. J., Zurek D. R. 2012, MNRAS, 423, 2
\bibitem[\protect\citeauthoryear{Maeder}{2009}]{maeder09} Maeder A. 2009,
  Physics, Formation and Evolution of Rotating Stars. Berlin: Springer-Verlag
\bibitem[\protect\citeauthoryear{Marks, Kroupa \& Baumgardt}{2008}]{marks08} Marks M.,
Kroupa P., Baumgardt H. 2008, MNRAS, 386, 2047
\bibitem[\protect\citeauthoryear{Marks \& Kroupa}{2010}]{marks10} Marks M.,
Kroupa P. 2010, MNRAS, 406, 2000
\bibitem[\protect\citeauthoryear{McKee \& Ostriker}{2007}]{mckee07} McKee C. F.,
Ostriker E. C. 2007, ARA\&A, 45, 565
\bibitem[\protect\citeauthoryear{Moody \& Sigurdsson}{2009}]{moody09}
Moody K., Sigurdsson S. 2009, ApJ, 690, 1370
\bibitem[\protect\citeauthoryear{Paczynsky \& Wiita}{1980}]{paczynsky80} Paczynsky B.,
Wiita P. J. 1980, A\&A, 88, 23
\bibitem[\protect\citeauthoryear{Park \& Ricotti}{2013}]{park13} Park K., Ricotti M. 2013, 
ApJ, submitted (arXiv:1211.0542)
\bibitem[\protect\citeauthoryear{Phinney \& Sigurdsson}{1991}]{phinney91} Phinney S. E.,
Sigurdsson S. 1991, Nature, 349, 220
\bibitem[\protect\citeauthoryear{Piotto et al.}{2007}]{piotto07}
Piotto G., Bedin L. R., Anderson J., King I. R., Cassisi S., Milone A. P., Villanova S., Pietrin-
ferni A., Renzini A. 2007, ApJ, 661, L53
\bibitem[\protect\citeauthoryear{Portegies Zwart et al.}{2004}]{portegieszwart04} Portegies Zwart S. F., 
Baumgardt H., Hut P., Makino J., McMillan S. L. W. 2004, Nature, 428, 724
\bibitem[\protect\citeauthoryear{Ruffert}{1994}]{ruffert94} Ruffert M. 1994, ApJ, 427, 342
\bibitem[\protect\citeauthoryear{Ruffert}{1997}]{ruffert97} Ruffert M. 1997, A\&A, 317, 793
\bibitem[\protect\citeauthoryear{Rybicki \& Lightman}{1979}]{rybicki79} Rybicki G. B.,
Lightman A. P. 1979, Radiative Processes in Astrophysics (New York: Wiley-Interscience)
\bibitem[\protect\citeauthoryear{Spitzer}{1969}]{spitzer69} Spitzer L. Jr. 1969, ApJ, 
158, 139
\bibitem[\protect\citeauthoryear{Spitzer}{1987}]{spitzer87} Spitzer L. Jr. 1987,
Dynamical Evolution of Globular Clusters (Princeton, NJ: Princeton Univ. Press)
\bibitem[\protect\citeauthoryear{Strader et al.}{2012}]{strader12} Strader J., Chomiuk L.,
Maccarone T. J., Miller-Jones J. C. A., Seth A. C. 2012, Nature, 490, 71
\bibitem[\protect\citeauthoryear{Tremaine, Ostriker \& Spitzer}{1975}]{tremaine75}
Tremaine S. D., Ostriker J. P., Spitzer L. Jr. 1975, ApJ, 196, 407
\bibitem[\protect\citeauthoryear{Tutukov}{1978}]{tutukov78} Tutukov A. V.
1978, A\&A, 70, 57
\bibitem[\protect\citeauthoryear{Vesperini \& Heggie}{1997}]{vesperini97}
Vesperini E., Heggie D. C. 1997, MNRAS, 289, 898
\bibitem[\protect\citeauthoryear{Vishniac}{1978}]{vishniac78} Vishniac E. T. 1978, ApJ, 223, 986
\bibitem[\protect\citeauthoryear{von Hippel \&
    Sarajedini}{1998}]{vonhippel98} von Hippel T., Sarajedini A. 1998,
  AJ, 116, 1789
\bibitem[\protect\citeauthoryear{Webb, Harris \& Sills}{2012}]{webb12} Webb J. J., 
Harris W. E., Sills A. 2012, ApJ, 759, 39
\bibitem[\protect\citeauthoryear{Yuan, Xie \& Ostriker}{2009}]{yuan09} Yuan F., 
Xie F., Ostriker J. P. 2009, ApJ, 691, 98
\bibitem[\protect\citeauthoryear{Zonoozi et al.}{2011}]{zonoozi11} Zonoozi A. H.,
Kupper A. H. W., Baumgardt H., Haghi H., Kroupa P., Hilker M. 2011,
MNRAS, 411, 1989
\bibitem[\protect\citeauthoryear{Zhang \& Fall}{1999}]{zhang99} Zhang
  Q., Fall S. M. 1999, ApJ, 527, 81

\end{thebibliography}
\end{document}